\documentclass[12pt,preprint]{aastex}
\def\etal{{et al.}}
\def\meszaros{M\'{e}sz\'{a}ros}

\begin{document}

\title{Testing Gamma-Ray Burst Jet Structure with the Distribution of Gamma-Ray Energy Release}
\author{L. Xu, X. F. Wu, and Z. G. Dai}
\affil{Department of Astronomy, Nanjing University, Nanjing
210093, China}

\begin{abstract}
We present a general method for testing gamma-ray burst (GRB) jet
structure and carry out a comprehensive analysis about the
prevalent jet structure models. According to the jet angular
energy distribution, we can not only derive the expected
distribution of the GRB isotropic-equivalent energy release for
any possible jet structure, but also obtain a two-dimensional
distribution including redshift $z$. By using the
Kolmogorov-Smirnov test we compare the predicted distribution with
the observed sample, and find that the power-law structured jet
model is most consistent with the current sample and that the
uniform jet model is also plausible. However, this conclusion is
tentative because of the small size and the inhomogeneity of this
sample. Future observations (e.g., {\em Swift}) will provide a
larger and less biased sample for us to make a robust conclusion
by using the procedure proposed in this paper.
\end{abstract}

\section{INTRODUCTION}
\label{sec:introduction} Growing evidence such as the achromatic
breaks in afterglow light curves of $\sim 20$ gamma-ray bursts
(GRBs) and the observation of polarized emission (Covino et al.
1999; Wijers et al. 1999) suggests that GRBs are produced by
collimated jets. Numerous models for the jet configuration have
been proposed due to its importance in understanding the burst
progenitor and GRB event rates. One leading model is the uniform
jet one (e.g., Rhoads 1997; Sari, Piran, \& Halpern 1999; Frail et
al. 2001; Dai \& Cheng 2001 ), in which the angular energy
distribution within the jet is uniform while the nearly constant
total energy is collimated into different opening angles. Another
is the structured jet model, in which the jet energy and structure
are approximately identical for all GRBs, but the energy per solid
angle $\epsilon(\theta)$ varies as a function of angle from the
jet axis within the structured jet (\meszaros, Rees, \& Wijers
1998; Dai \& Gou 2001; Rossi, Lazzati, \& Rees 2002; Zhang \&
\meszaros~2002). Furthermore, although the specific distribution
of jet opening angles in the uniform jet model is uncertain,
several kinds of distribution such as the power-law, Gaussian and
exponential function have been suggested. In this paper we discuss
these possible models specifically.

 Since a few leading models of jet structure can all
explain the observed achromatic breaks in afterglow light curves,
some authors attempted to test the geometrical configuration by
using other observations. For example, the jet structure could be
constrained  by considering the luminosity function (e.g., Firmani
et al. 2004; Lloyd-Ronning, Dai, \& Zhang 2004; Guetta, Piran, \&
Waxman 2005). The $\log N - \log P$ plot was also used to constrain
the jet opening angle distribution for the uniform jet and the star
forming rate (Lin et al. 2004; Guetta, Piran, \& Waxman 2005).
Perna, Sari, \& Frail (2003) predicted the observed one-dimensional
(1D) distribution of viewing angles, $n(\theta)=dn/d\theta$, in the
power-law structured jet model. They found that the predicted
distribution could fit the current sample of 16 bursts with modelled
angles and known redshifts. Considering the same model, Nakar,
Granot, \& Guetta (2004) carried out an analysis about the
two-dimensional (2D) distribution $n(z,\theta)=d^2n/dzd\theta$.
However, they found that the 2D prediction shows a very poor
agreement with the same sample, and the hypothesis that the data are
drawn from this model is rejected at the $99\%$ confidence level by
a 2D Kolmogorov-Smirnov (K-S) test. Based on the standard energy
reservoir of GRB jets and the relationship between the cosmic rest
frame GRB spectral peak energy and the isotropic gamma-ray energy,
i.e., $E_{\rm peak} \propto E_{\rm iso}^{1/2}$, Liang, Wu, \& Dai
(2004) simulated a sample including $10^6$ GRBs to test the GRB
probability distribution as a function of viewing angle and
redshift, and supported the power-law structured jet model. Dai \&
Zhang (2005) summarized the above tests and performed a global test
on the quasi-universal Gaussian-like structured jet. Together with
their previous tests with the observed jet break angle vs. isotropic
energy $(E_{\rm iso}-\theta_j)$ and observed peak energy vs. fluence
relations ($E_{\rm peak}^{\rm obs}$ vs. fluence), Zhang et al.
(2004) concluded that the current GRB data are generally consistent
with the Gaussian-like jet model. These tests can only tentatively
support certain possible configuration of the GRB jet, partly
because the size of the sample use in the test is too small (but see
Cui, Liang, \& Lu 2005, who recently adopted the BATSE sample). In
this paper, we propose a new method to test the GRB jet structure by
using the sample of bursts with measured fluences and redshifts. By
considering three jet structure models, i.e., the power-law
structured jet model, the Gaussian structured jet model, and the
uniform jet model, we work out the theoretical distribution of burst
isotropic-equivalent energy release in both 1D distribution
$\dot{N}(E)=dN/dE$ and 2D distribution $\dot{N}(E,z)=d^2N/dEdz$
(hereafter, we use $E$ instead of $E_{\rm {iso}}$ for brevity), and
compare them with the observed distribution of $E$ derived from the
sample with measured fluences and redshifts.

 This paper is organized as follows:
 In \S \ref{sec:models} we present several prevalent models for jet structure
 and the probability distribution of the isotropic-equivalent energy under these
 models, while in \S \ref{sec:detections}
 we describe our methodology in detail.
 In \S \ref{sec:result} we discuss what each model predicts
 for the observed distribution (in 1D and 2D) of GRB isotropic-equivalent
 energy release and assess how our numerical
 results are compared with the current sample by using K-S test.
 We summarize our results and draw our main conclusions in \S
 \ref{sec:discussion}.

\section{THEORY}
\label{sec:theory}
\subsection{Models for jet structure}
\label{sec:models}

Since the energy release in an explosion plays an important role
in determining the progenitors of GRBs and the physics of their
central engine, some authors have carried out plenty of analyses
about the gamma-ray energy release distribution. Within the
uniform jet model, Frail et al. (2001) found that the gamma-ray
energy release after beaming correction is narrowly clustered
around $5 \times 10^{50}$ ergs by fitting a sample of 16 well
observed GRBs. Using a larger sample, Bloom, Frail, \& Kulkarni
(2003) obtained a similar result that the gamma-ray energies are
well clustered around a median value of $1.33 \times 10^{51}$
ergs. The gamma-ray energy release should become larger if the
gamma-ray radiation efficiency is considered (Xu \& Dai 2004).
Panaitescu \& Kumar (2001, 2002) modelled the broadband emission
of several GRB afterglows and obtained a similar value of the
kinetic energy. Furthermore, Wu, Dai, \& Liang (2004) derived a
jet break time-flux density relationship and constrained some
physical parameters of gamma-ray burst afterglows. The jet break
relation supports the ``standard candle" hypothesis of the
afterglows by an entirely different approach. We therefore adopt a
standard value for GRB jet energy $E_{j}=10^{51}$ ergs throughout
this paper. According to the presently measured
isotropic-equivalent energy, we adopt the upper limit of the
isotropic-equivalent energy as $E_{\rm max}=10^{55}$ ergs.

According to the angular energy distribution for any plausible jet
model, one can obtain the isotropic-equivalent energy as a
function of $\theta$. The isotropic-equivalent energy, $E$, is
defined as $E(\theta)=4\pi \epsilon(\theta)$, where $\theta$ is
the angle from the jet axis and $\epsilon(\theta)$ is the energy
per solid angle.

In the power-law structured jet (PLSJ) model, all GRBs are supposed to have a
similar jet profile, and their different observed properties are due to
different observer's viewing angles. Under the power-law jet scenario, the
angular distribution of the isotropic-equivalent energy is
\begin{equation}
E(\theta)=\frac {E_{c}} {[1+(\theta/\theta_{c})^{k_{E}l_{\rm
E}}]^{1/l_{E}}},~~~~~~(0\leq\theta \leq \pi/2),
\label{eqn:plj-structure}
\end{equation}
where $l_{E}$ determines the sharpness of the transition from the
core to wing, and $k_{E}$ represents the profile of the jet wing.
The opening angle of the core $\theta_c$ is introduced to avoid
the divergence of the total energy. $E_{c}$ is defined as
$E_{c}=E(0)$, and equals to the maximum isotropic-equivalent
energy $E_{\rm max}$. Below we assume $l_E=+\infty$ for
simplicity. If all bursts were observable, then in a structured
jet model, the probability for a viewing angle $\theta$ is
$P(\theta)d{\theta}\approx \theta d\theta$. From the relation
$P(\theta)d\theta=P(E)dE$, we obtain the normalized probability
density at a given isotropic-equivalent energy $E$,
\begin{equation}
P(E)\approx\frac {8}{\pi^2} \frac {\theta_c^2}{k_EE_c} \left(
\frac {E_c}{E} \right)^{1+2/k_E}, ~~~( ( {2\theta_c}/{\pi})^{k_E}
E_{c}\leq E\leq E_{c}).\label{eqn:pl-pro}
\end{equation}

In the Gaussian structured jet (GSJ) model, the
isotropic-equivalent energy distribution that varies with the
angle from the jet axis is (Zhang \& \meszaros~2002)
\begin{equation}
E(\theta)=E_{c} e^{-\frac {\theta^2}{2\theta_c^2}},~~~~~~(0\leq
\theta\leq \pi/2), \label{eqn:gj-structure}
\end{equation}
where $E_{c}=E(0)$, and $\theta_c$ is the characteristic jet
angular width. According to the same derivation as in the
power-law structured jet model, we have the normalized probability
at a given isotropic-equivalent energy $E$,

\begin{equation}
P(E)={\frac {8}{\pi^2}}{\frac {{\theta_c}^2}{E},}~~~({E_{c}}
e^{-\pi^2/8{\theta_c}^2}\leq E\leq E_{c}). \label{eqn:gj-pro}
\end{equation}
Similarly we assume $E_{c}=10^{55}$ ergs.

In the uniform jet (UJ) model (e.g., Lamb, Donaghy, \& Graziani 2005), all GRBs
produce jets with different jet opening angles $\theta_{j}$. The structure of a
uniform jet reads
\begin{equation}
 E(\theta)=\left\lbrace \begin{array}{ll} E, ~~~~~~0\leq \theta\leq \theta_{j},\\
                                         0,~~~~~~~\theta_{j}< \theta \leq \pi/2.
                       \end{array} \right.
\label{eqn:uj-structure}
\end{equation}
Assuming $f(E)$ is the probability that a GRB releases the
isotropic energy between $E$ and $E+dE$, the probability that we
observe a specified $E$ is given by $P(E)=f(E)(1-\cos
\theta_{j})$. Using the Frail's relation, $E_{j}= E (1-\cos
\theta_{j})\approx {\rm constant}$, we obtain the probability
distribution of the isotropic energy

\begin{equation}
P(E) \approx f(E){\frac {E_{j}}{E}}, ~~~~(E_{j}\leq E\leq
E_{{max}}),\label{eqn:uj-pro}
\end{equation}

Next we turn to discuss the specified form of $f(E)$. As for the
GRB opening angle distribution, Lin et al. (2004) studied three
cases in which the GRB opening angle $\theta_{j}$ follows a
power-law, exponential, or Gaussian distribution. Here we also
consider these three cases. We assume that the first probability
distribution of isotropic-equivalent energy follows a power-law
function with index $\zeta$,

\begin{equation}
f(E)=\frac {1+\zeta}{E_{\rm {max}}^{1+\zeta}-E_{
j}^{1+\zeta}}E^{\zeta},~~~(E_{j}\leq E\leq E_{\rm {max}}).
\label{eqn:uj-pl}
\end{equation}
It is worth noting that $f(E)$ should be normalized, i.e.
$\int_{E_{j}}^{E_{\rm {max}}}f(E)dE=1$. The second probability
distribution for $f(E)$ is an exponential function with parameter
$\lambda$,
\begin{equation}
f(E)=\frac {\lambda e^{\lambda E_{j}}}{1-e^{-\lambda(E_{\rm
{max}}-E_{j})}}~e^{-\lambda E},~~~(E_{j}\leq E\leq E_{\rm {max}}),
\label{eqn:uj-exp}
\end{equation}
which has also been normalized. Finally, if $f(E)$ is a Gaussian
distribution with the mean energy $\bar{E}$ and standard scatter
$\sigma$, we have
\begin{equation}
f(E)=\frac {a}{\sqrt{2\pi}\sigma}e^{-\frac
{(E-\bar{E})^2}{2\sigma^2}},~~~(E_{j}\leq E\leq E_{\rm {max}}),
\label{eqn:uj-gs}
\end{equation}
where $a$ is the normalization factor satisfying $\int_{E_{
j}}^{E_{\rm {max}}}f(E)dE=1$.

\subsection{Detections}
\label{sec:detections}

For a power-law photon spectrum, the photon luminosity in the
triggering band $\nu_{l}-\nu_{u}$ is given by
\begin{equation}
L_{{\rm ph},[\nu_{l},\nu_{u}]}=\frac {E} {T h \nu_{l}} \frac
{\alpha -2}{\alpha - 1}\left(\frac {\nu_{l}}{\nu_1} \right
)^{2-\alpha} \frac {1-\left(\nu_{u}/\nu_{l} \right )^{1-\alpha}}
{1-\left({\nu_2}/{\nu_1} \right )^{2-\alpha}},
\label{eqn:photon-L}
\end{equation}
where $\alpha$ is the photon spectrum index, and $\nu_{l}$ and $\nu_{u}$ are
the lower and upper limiting frequencies of a GRB detector respectively. For
the BATSE triggering channel, $\nu_{l}=50~{{\rm keV}}$ and $\nu_{u}=300~{\rm
{keV}}$. The $E$ is the isotropic-equivalent $\gamma$-ray energy in a range of
$\nu_1=0.1$ keV to $\nu_2=10^4$ keV at the GRB's cosmological rest frame. The
mean spectral peak energy for GRBs is $\langle E_{p}\rangle\sim 250~{\rm
{keV}}$ (Preece et al. 2000). According to the spectral fit, we take $\alpha
\approx 1$ for the frequency range $\nu_{l}<\nu<\nu_{u}$ in the following
calculations (Band et al. 1993). Then the photon luminosity can be simplified
as
\begin{equation}
L_{{\rm ph},[\nu_{l},\nu_{u}]}=\frac {E} {T h(\nu_2-\nu_1)} \ln
\left (\frac {\nu_{u}}{\nu_{l}} \right), \label{eqn:photon-L1}
\end{equation}
where $T$ is an ``effective" duration that the burst would have if
its energy output is constant at the peak value rather than the
highly variable one.

The detector can be triggered if the photon flux is greater than the limiting
photon flux that is determined by the threshold of the GRB detector. Thus we
can determine the maximum redshift $z_{\rm {max}}$ of the burst through
\begin{equation}
\frac {L_{{\rm {ph}},[\nu_{l},\nu_{u}]}(E,T)}{4\pi D^2(z_{\rm
{max} })(1+z_{\rm {max }})^\alpha}= F_{\rm ph,lim},
\label{eqn:zmax}
\end{equation}
where $D(z)$ is the comoving distance of the source to the
observer and $F_{\rm ph, lim}$ is the triggering threshold of the
detector.

The observed burst rate with isotropic-equivalent energy between
$E$ and $E + d E$ is given by
\begin{equation}
\frac {d^2N(E)} {dt_{\rm obs}~ d{E}}=P(E)\int_{0}^{z_{\rm
max}(E,T)} \frac {R_{\rm GRB}}{1+z} \frac {dV}{dz}dz,
\label{eqn:birthrate}
\end{equation}
where $R_{\rm GRB}$ is the GRB rate per unit comoving volume per
unit comoving time, and $V(z)$ is the comoving volume. In a flat
cosmology $dV(z)/dz=4\pi D^2(z)~d D(z)/dz$.

Nakar et al. (2004) used the flux table of the BATSE 4B Catalog
and Band spectrum to estimate the distribution of $T$. They found
a lognormal distribution,
\begin{equation} \frac {d\bar{P}}{d\ln T}=T \bar{P}(T)=\frac
{1}{\sigma_{\ln T}\sqrt{2\pi} }\exp \left[{-\frac {(\ln
T-\mu)^2}{2{\sigma}_{\ln T}^2}}\right ],
\label{eqn:T-distribution}
\end{equation}
where $\mu=2.15$ and $\sigma_{\ln
T}=0.87~(T=8.6_{-5}^{+12}{\rm~s})$.

Next, we consider the selection effect in redshift identification.
Bloom (2003) found that there are strong observational biases in
ground-based redshift discovery, and suggested to use a
probability function related to the luminosity distance to
minimize this effect. $P_{L}(z)$ is constructed:
\begin{equation}
P_{L}(z)=\left\lbrace \begin{array}{ll} 1, ~~~~~~~~~~~~~~~~~z\leq z_{l},\\
                                         \left ( \frac {D_{L}(z)}{D_{L}(z_{l})} \right )^{L},  ~~~~~~z >z_l.
                       \end{array} \right.
\label{eqn:redshift-selection}
\end{equation}
Via K-S test, Bloom constrained  $z_{l}<1.25$ and $-3<L<-1$. We
here adopt the value of  $z_{l}=1$ and allow the parameter $L$ to
vary between $-3$ and $ -1$ case-by-case in order to get a better
fit. Using the distribution of $T$ and taking into account the
redshift selection effect, we generalize the distribution of
isotropic-equivalent jet energy as
\begin{equation}
\dot{N}(E)=\frac {d^2N(E)} {dt_{\rm obs}~
d{E}}=P(E)\int_{0}^{+\infty}\bar{P}(T)dT \int_{0}^{z_{\rm
{max}}(E,T)} P_{L}(z)\frac {R_{\rm GRB}}{1+z} \frac {dV}{dz}dz.
\label{eqn:one-d}
\end{equation}
Furthermore, since in the 1D analysis the integration  over
redshift hides certain important information about the
distribution as a function of $z$, we extend the 1D distribution
$\dot{N}(E)$ to the 2D distribution $\dot{N}(E,z)$,

\begin{equation}
\dot{N}(E,z)=\frac {d^3N(E)} {dt_{\rm obs}~ d{E}~dz}=P(E)\left (
\int_{0}^{T_{\rm {max}}(E,z)} \bar{P}(T)dT \right )P_{L}(z) \frac
{R_{\rm GRB}}{1+z} \frac {dV}{dz}, \label{eqn:two-d}
\end{equation}
where $T_{\rm {max}}(E,z)$ is determined by equation
(\ref{eqn:zmax}).

Finally, we consider the GRB event rate. It has been suggested
that GRBs follow the star formation rate, because GRBs are
probably produced in the final gravitational collapse of massive
stars (e.g. Woosley 1993; Paczynski 1998). Here we assume that the
rate of GRBs traces the global star formation history of the
universe, $R_{\rm {GRB}}(z)\propto R_{\rm SF}(z) \propto R_{\rm
SN}(z)$. Following Porciani \& Madau (2001), we employed three
different kinds of global star formation rate (SFR) model. These
three SFR models are similar at $z<1$. The main difference is at
high redshifts: in model 1, the SFR decreases at $z>1.5$; in model
2, the SFR contains the dust extinction effect and remains
constant at $z>2$; and in model 3, the SFR increase at high
redshifts after a correction due to a large amount of dust
extinction. These SFR models are all in an Einstein-de Sitter
universe with parameter $\Omega_{M}=1.0,~\Omega_{K}=0.0,~
\Omega_{\Lambda}=0.0, $ and $H_0=65~{\rm {km~s^{-1}~Mpc^{-1}}}$.
We can obtain the SFR in any cosmology with parameters
$\Omega_{M},~\Omega_{\Lambda},~\Omega_{K}$ and $H_0$ through the
formulation in the appendix of Porciani \& Madau (2001).

We assume that all stars with masses $M>30M_\odot$ explode as Type
Ib/Ic SNe, and adopt the initial mass function suggested by Madau
$\&$ Pozzetti (2000). The SN Ib/Ic rate can then be estimated by
$R_{\rm {SN~Ib/c}}=f_{\rm {SN~Ib/c}}R_{\rm {SF}}=1.8\times 10^{-3}
M_\odot^{-1}R_{\rm {SF}}$. Lamb (1999) showed that the observed
ratio of the rate of GRBs to the rate of Type Ib/Ic SNe in the
observable universe is ${R_{\rm {GRB}}}/{R_{\rm {Type~ Ib/c}}}
\sim 10^{-5}$. Therefore the GRB event rate can be approximated by
$R_{\rm {GRB}}=f_{\rm {GRB}}R_{\rm {SF}}=10^{-8}
M_\odot^{-1}R_{\rm SF}$.

\section{Results}
\label{sec:result}
In order to test the configuration of jets, we firstly perform a
theoretical analysis in two limits: the rate of GRBs at low and
high values of isotropic-equivalent energy. The analysis can also
predict and test our numerical results. Secondly, we present
detailed results of  numerical calculations, which are well
consistent with the theoretical one.

We used the same parameters as in Perna et al. (2003) and Nakar et
al.(2004). The triggering threshold is $F_{\rm
{ph,lim}}=0.424~{\rm photons~cm^{-2}~s^{-1}} $, and the
cosmological parameters are:
$\Omega_{M}=0.3,~\Omega_{\Lambda}=0.7,$ and $H_0=71~{\rm
km~s^{-1}~Mpc^{-1}}$.

\subsection{Theoretical analysis}
\label{sec:theoretical-ana}

The comoving distance in the currently adopted cosmology can be
approximated by (Wu, Dai, \& Liang 2004)
\begin{equation}
D(z)\simeq \frac {c}{H_0} \frac {z}{1+0.29z}.
\label{eqn:distance}
\end{equation}
Combined with equation (\ref{eqn:zmax}), the photon flux in both
limits reads
\begin{equation}
F_{\rm {ph,lim}}\approx 0.55~{\rm {photons}\cdot{cm}^{-2}\cdot{s}^{-1}} E_{53}
\left ( \frac {T}{8~\rm s }\right )^{-1}\times \left\lbrace
\begin{array}{ll} z_{\rm {max}}^{-2}~, ~~~~~~~~~z_{\rm {max}}<1,\\
                                         0.29^2z_{\rm {max}}^{-1}~,  ~~~z_{\rm
                                         {max}}>4.
                       \end{array} \right.
\label{eqn:photon-phlux}
\end{equation}
We take the conventional notation $Q=Q_{x}\cdot 10^x$. For a given
triggering instrument, $F_{\rm {ph,lim}}$ is known. Therefore in
the limit of $z_{\rm {max}}<1$ the maximum redshift that a GRB can
be detected is
\begin{equation}
z_{\rm {max}}=1.15 \left ( \frac {F_{\rm {ph,lim}}}{F_{\rm
{ph,lim}}^{\rm {BATSE}}}\right )^{-1/2} E_{53}^{1/2}\left ( \frac
{T}{8~\rm s }\right )^{-1/2}, \label{eqn:limit1}
\end{equation}
while in the limit of $z_{\rm {max}}>4$, it follows
\begin{equation}
z_{\rm {max}}=0.11 \left ( \frac {F_{\rm {ph,lim}}}{F_{\rm
{ph,lim}}^{\rm {BATSE}}}\right )^{-1} E_{53}\left ( \frac
{T}{8~\rm s }\right )^{-1}. \label{eqn:limit2}
\end{equation}
Here we scale the detection threshold to that of the BATSE
triggering channel. Assuming an empirical $T \sim 8~\rm s $, the
maximum redshift is $z \propto E^{1/2}$ for low energies
$E<10^{53}~{\rm {ergs}}$, and $z \propto E$ for extremely large
energies $E > 10^{54}~{\rm {ergs}}$. In the following analysis, we
take the star formation rate model 2 as an example.

In the case of $z_{\rm {max}}<1$, i.e., for GRBs with low isotropic energy of
$E<10^{53}~\rm ergs$, the integration over redshift $z$ in equation
(\ref{eqn:birthrate}) gives $43 h_{65} f_{\rm {GRB},-8}({F_{\rm
{ph,lim}}}/{F_{\rm {ph,lim}}^{\rm {BATSE}}})^{-3/2} E_{53}^{3/2} ( {T}/{8~\rm s
})^{-3/2}~{\rm yr^{-1}}$, where $h_{65}=H_0/65{~\rm km \cdot s^{-1} \cdot
Mpc^{-1}}$. The differential rate of bursts for the power-law structured jet
model is
\begin{equation}
\frac {d^2N(E)} {dt_{\rm {obs}}~ d{\log_{10} E}}\simeq 0.4 h_{65}\left(\frac
{k_E}{2.0} \right)^{-1} \left(\frac {\theta_c}{0.01}\right)^2
E_{c,55}^{2/k_E}f_{\rm {GRB},-8} \left ( \frac {F_{\rm {ph,lim}}}{F_{\rm
{ph,lim}}^{\rm {BATSE}}}\right )^{-3/2}\left ( \frac {T}{8~\rm s }\right
)^{-3/2} E_{53}^{1/2}~{\rm yr^{-1}}. \label{eqn:pl-limit1}
\end{equation}
Similarly the differential rate for the Gaussian structured jet
model is
\begin{equation}
\frac {d^2N(E)} {dt_{\rm {obs}}~ d{\log_{10} E}}\simeq 0.2
h_{65}\left ( \frac {\theta_{c}}{0.05}\right )^2 f_{\rm {GRB},-8}
\left ( \frac {F_{\rm {ph,lim}}}{F_{\rm {ph,lim}}^{\rm
{BATSE}}}\right )^{-3/2}\left ( \frac {T}{8~\rm s }\right )^{-3/2}
E_{53}^{3/2}~{\rm yr^{-1}}. \label{eqn:gj-limit1}
\end{equation}
For the uniform jet model, we just present the rate in the case
that the distribution of isotropic-equivalent energy is an
exponential function,
\begin{equation}
\frac {d^2N(E)} {dt_{\rm {obs}}~ d{\log_{10} E}}\simeq
14h_{65}\lambda_{-53} E_{j,51}^{1/2} (
{e^{0.01\lambda_{-53}E_{j,51 }}} ) f_{\rm {GRB},-8} \left ( \frac
{F_{\rm {ph,lim}}}{F_{\rm {ph,lim}}^{\rm {BATSE}}}\right
)^{-3/2}\left ( \frac {T}{8~\rm s }\right
)^{-3/2}E_{53}^2e^{-\lambda_{-53} E_{53}}~{\rm yr^{-1}},
\label{eqn:uj-exponential-limit1}
\end{equation}
where $\lambda=10^{-53}\lambda_{-53}~{\rm erg}^{-1}$ and $E_{{j},51}$ is the
jet energy in units of $10^{51}$ ergs.

In the case of $z_{\rm {max}}>4$, i.e., $E > 10^{54}$ ergs, the integration
over redshift $z$ in equation (\ref{eqn:birthrate}) gives $\sim 1.86 \times
10^{3}h^4_{65} f_{\rm {GRB},-8}g(z_{\rm {max}})~{\rm yr^{-1}}$. The value of
$g(z_{\rm {max}})$ ranges from $0.03$ to $0.28$. To obtain an estimation, we
adopt $g(z_{\rm {max}})\approx 0.1$. Then the differential GRB rate for the
power-law structured jet model is
\begin{equation}
\frac {d^2N(E)} {dt_{\rm {obs}}~ d{\log_{10} E}}\simeq 1.7h^4_{65} \left (\frac
{k_E}{2.0}\right)^{-1}\left(\frac {\theta_c}{0.01}\right)^2 E_{c,55}^{2/k_E}
 f_{\rm {GRB},-8} \frac {g(z_{\rm {max}})}{0.1}
 E_{53}^{-1}~{\rm yr^{-1}}.
\label{eqn:pl-limit2}
\end{equation}
Similarly, the rate for the Gaussian structured jet model is
\begin{equation}
\frac {d^2N(E)} {dt_{\rm {obs}}~ d{\log_{10} E}}\simeq 0.85
h^4_{65} f_{\rm {GRB},-8}  \frac {g(z_{\rm {max}})}{0.1}\left (
\frac {\theta_{c}} {0.05}\right
 )^2~{\rm yr^{-1}}.
\label{eqn:gj-limit2}
\end{equation}
At last, for the uniform jet model, if the distribution of jet
isotropic-equivalent energy conforms to an exponential
distribution, then the differential burst rate is
\begin{equation} \frac
{d^2N(E)} {dt_{\rm {obs}}~ d{\log_{10} E}}\simeq 60.4 h^4_{65}
\lambda_{-53} ({e^{0.01\lambda_{-53}E_{j,51 }}}) E_{j,51}^{1/2}
f_{\rm {GRB},-8} \frac {g(z_{\rm
{max}})}{0.1}E_{53}^{1/2}e^{-\lambda_{-53} E_{53}} ~{\rm yr^{-1}}.
\label{eqn:gj-limit2}
\end{equation}

\begin{figure}
\plotone{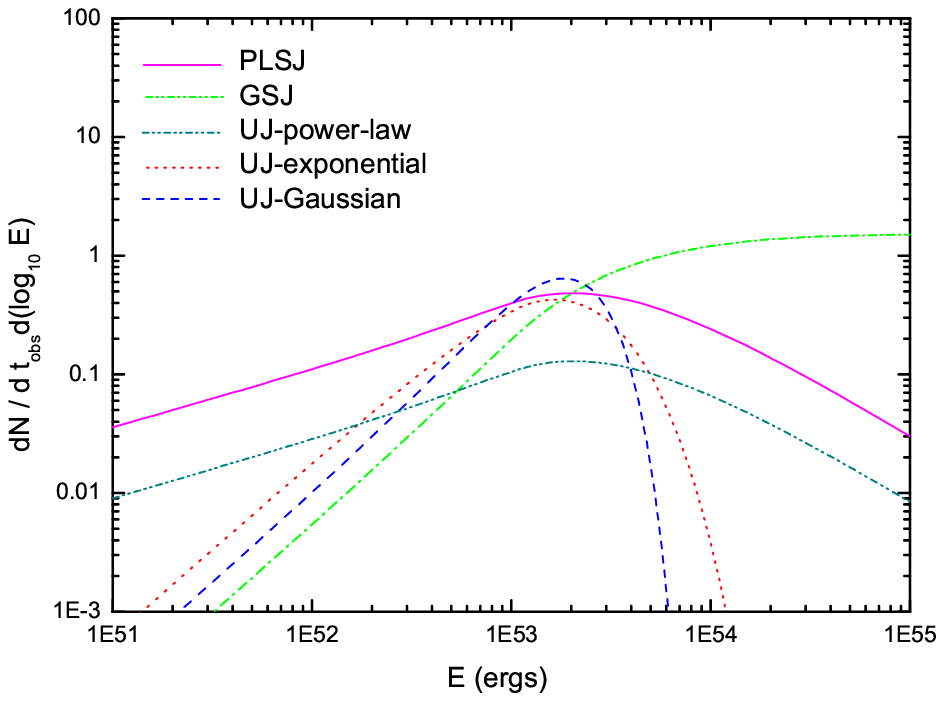} \caption{ The observed differential rate of GRBs as a
function of $E$ for 5 kinds of jet model. The parameter $T$ is chosen to be 8 s
for all GRBs. Other parameters are assumed: $L=-1$, $E_j=10^{51}~{\rm
ergs},~f_{\rm GRB}=10^{-8}~{M^{-1}_\odot}$, and $F_{\rm {ph,lim}}=0.424~{\rm
photons \cdot cm^{-2} \cdot s^{-1}}$.  For power-law structured jet models,
$k_E=2.0$ and $\theta_c=0.01$. For the Gaussian structured jet model,
$\theta_c=0.05$. For the uniform jet model $\lambda=0.7 \times 10^{-53}~{\rm
ergs^{-1}}$ (the exponential case); $\zeta=-1.1$ (the power-law case);
$\bar{E}=0.7\times 10^{53}~{\rm ergs}$, $\sigma=1.4\times 10^{53}~{\rm ergs}$
(the Gaussian case). Here we take the star formation rate model 2 as an
example. \label{fig:theoretical-ana}}
\end{figure}

Figure \ref{fig:theoretical-ana} shows differential rates of GRBs
determined by equation (\ref{eqn:birthrate}). Different lines
correspond to the predicted distributions derived from different
possible models. It can be seen that the above analytical
expressions describe the rate quite well at low and high limits of
$E$. The predicted GRB rates for these models differ from each
other significantly, especially at extremely low and high energy
limits. Such evident differences enable us to find the best model
for the jet structure, if the sample contains enough bursts.
Therefore the 1D comparison between the predicted distribution of
$E$ and the observed sample is available, although the integration
over redshift conceals much information.

\subsection{Numerical analysis}
\label{sec:numerical-ana}

To compare the theoretical distributions derived in \S
\ref{sec:theory}, we consider a sample of GRBs whose fluences and
 redshifts are measured. Such a sample including 41
 bursts are listed in Table 1. This table does not contain GRBs
 such as GRB 980329, 980519, 000630, 020331, 030115 and
 040511, because of the large uncertainties of their redshifts.
 We rule out 2 GRBs in this table: GRB $980425$ and $031203$,
 because they form a peculiar subclass characterized by
 their unusually low isotropic gamma-ray energy releases and other unusual
 properties. For example, these two GRBs violate the $E_{\rm \gamma, iso}-E_{\rm peak}$
 relation and the luminosity-spectral lag relation (Sazonov, Lutovinov, \& Sunyaev 2004).
Furthermore, Yamazaki, Yonetoku and Nakamura (2003) found that the
observed unusual properties of the prompt emission of GRB 980425
could be explained by using an off-axis jet model of GRBs, such as
the extremely low isotropic equivalent $\gamma$-ray energy, the low
peak energy, and the high fluence ratio, and the long spectral lag.
Ramirez-Ruiz et al. (2004) also argued that the observed data for
GRB 031203 are more consistent with a GRB seen at an angle of about
twice the opening angle of the central jet. Since we do not consider
the off-axis case in our model analysis, we rule out these two GRBs
in the numerical analysis. Therefore, we finally have 39 bursts
available in our analysis.  We have calibrated the
isotropic-equivalent energy release in a fixed cosmological
bandpass, 0.1 keV-10 MeV, using the cosmological k-correction
(Bloom, Frail, \& Sari 2001). The calorimetric isotropic-equivalent
energy is given by $E={4\pi D_{\rm L}^2}S_{[0.1-10^4]}/{(1+z)}$.

 We calculate the 1D distribution $\dot{N}(E)$
 under three jet structure scenarios, and perform K-S
 test to assess the compatibility of the theoretical distributions
 with the observed data. We describe the significance level for the result of K-S test as $P_{\rm k-s}$.
 Small values of $P_{\rm k-s}$ show that the theoretical distributions
 and the observed data are significantly different (Press et al.
 1997). In the analysis below, we take the star formation rate model 2 and $L=-1.0$ as
 an example. We also perform the test for other star formation rate models as well as
 other values of $L$, and ultimately obtain similar results.

\begin{figure}
\plotone{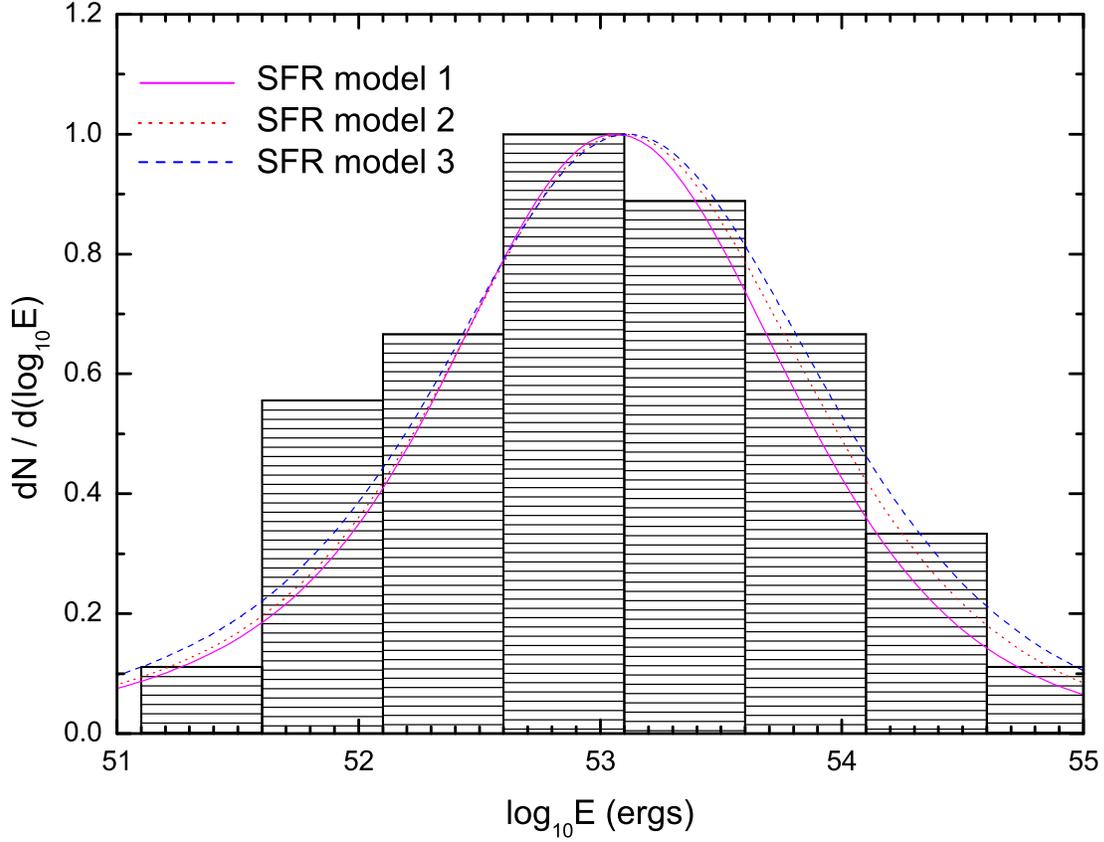} \caption{ Theoretical distribution of the
isotropic-equivalent energy in the power-law structured jet model
(lines), compared with the observed distribution from a sample of
39 GRBs with known redshifts detected so far (histogram).
Different lines correspond to different star formation rate
models. The parameters: $k_E=2.3$ and $\theta_c=0.02$. Here we
take $L=-1$ as an example.\label{fig:powerlaw1d}}
\end{figure}

\begin{figure}
\plotone{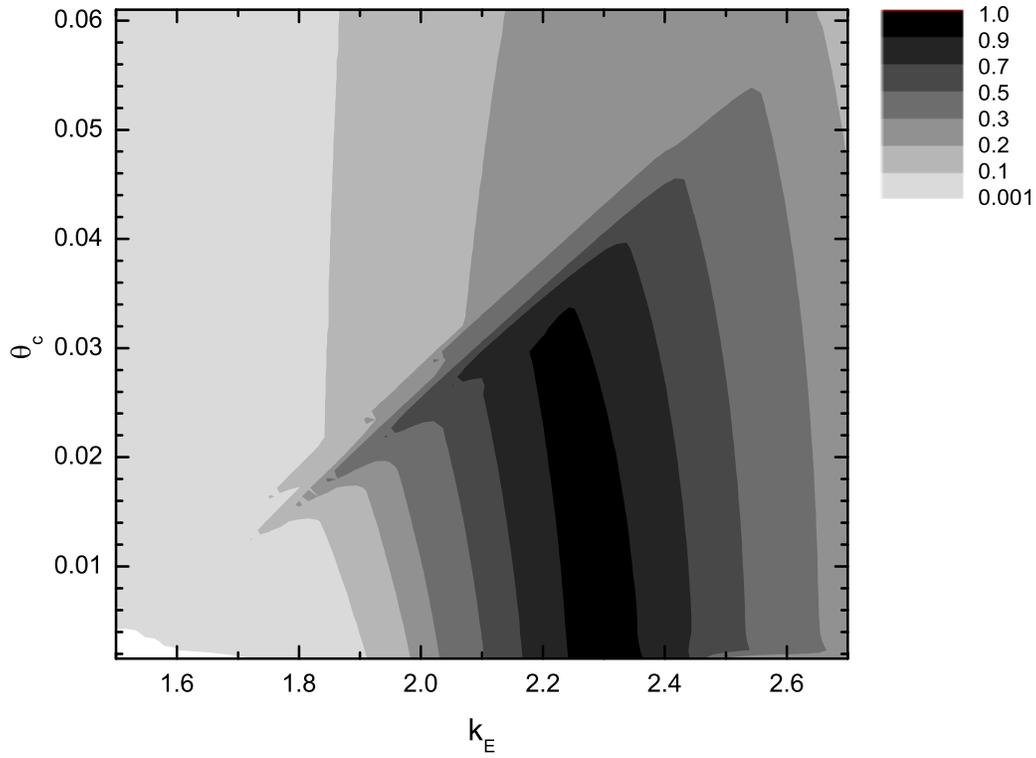} \caption{The Gray scale map of the K-S test result for
different values of $k_E$ and $\theta_c$. Every point in this plot corresponds
to a set of parameters for the PLSJ model and the color of the point represents
how well the PLSJ model with such parameters is compared with the observed
data. Here we take the star formation rate model 2 and $L=-1$ as an example.
\label{fig:powerlawks}}
\end{figure}

 Figure \ref{fig:powerlaw1d} shows that the predicted distribution peaks
 approximately at the same energy for different SFR models.
 Three SFR models result in slight differences in the
 predicted distribution if we adopt other parameters with the same values.
 In Figure \ref{fig:powerlawks} the 2D grey contours show the confidence level (for the null hypothesis
 that the observed data are drawn from this model) as a function of $k_E$ and $\theta_c$.
 We find that the predicted distribution is consistent with the observations at the confidence level of
 $>40\%$ while the parameters $k_E$ and $\theta_c$ are changed in a wide range.
 For other star formation rate models, the power-law structured jet
 model also agrees with the observational data reasonably well.

\begin{figure}
\plotone{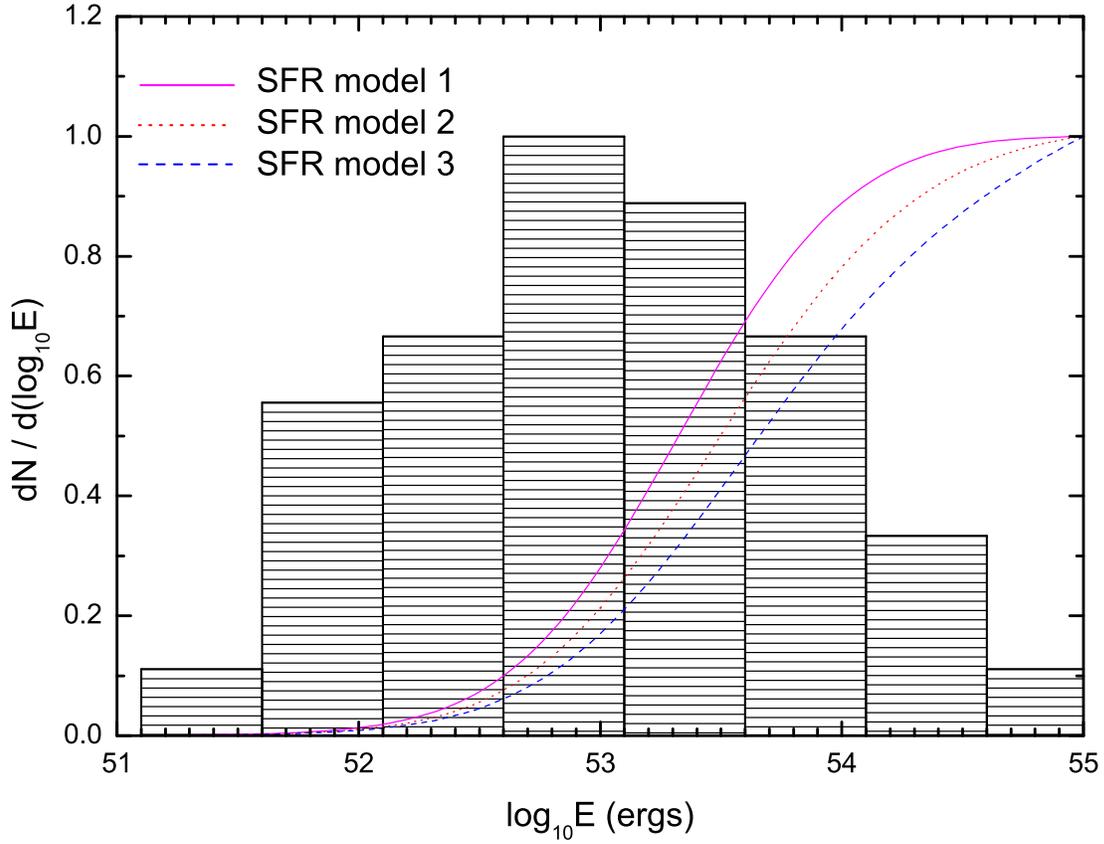} \caption{ Theoretical distribution of the
isotropic-equivalent energy in the Gaussian structured jet model
(lines), compared with the observed distribution from a sample of
39 GRBs with known fluences and redshifts detected so far
(histogram). Different lines correspond to different star
formation rate models. The parameter $\theta_c=0.01$. Here we take
$L=-1$ as an example. \label{fig:gaussian1d}}
\end{figure}

\begin{figure}
\plotone{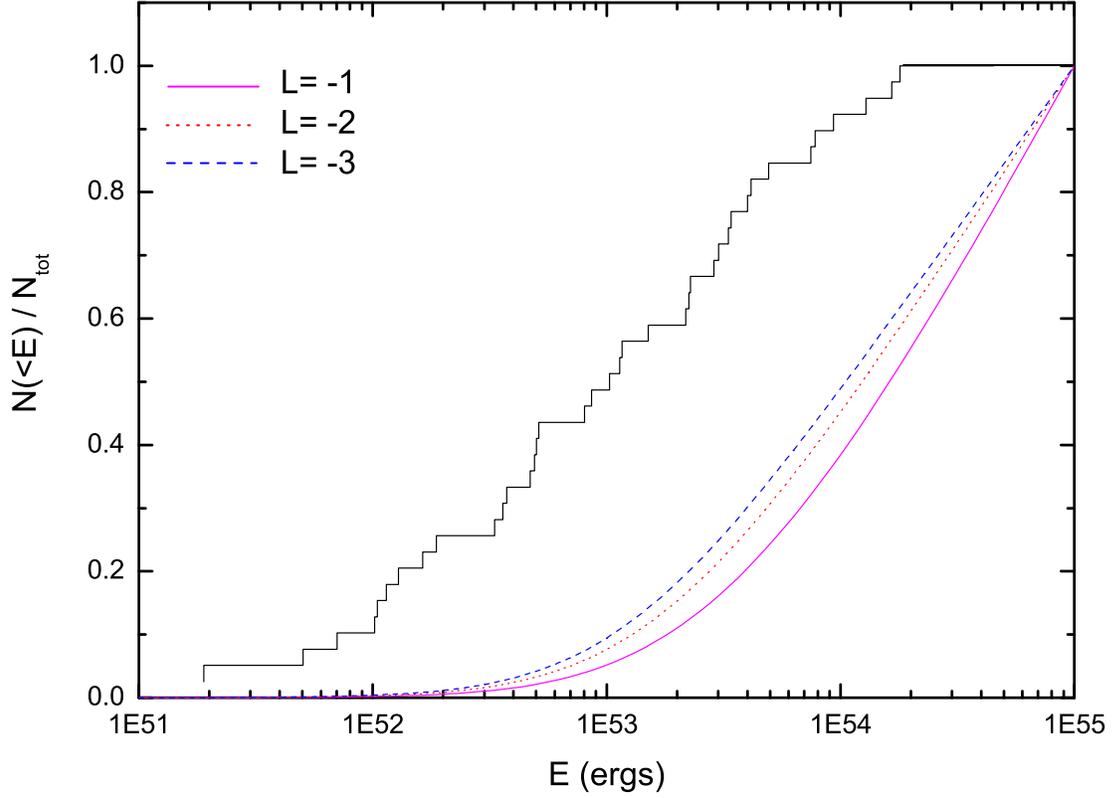} \caption{Cumulative distribution for the
Gaussian structured jet model. Three values of $L$, the index in
the redshift-identification selection effect, are considered. The
cumulative histogram is plotted from the sample. The parameter
$\theta_c=0.01$. Here we take star formation rate model 2 as an
example. \label{fig:gaussianks}}
\end{figure}

  However, the distribution derived from the Gaussian jet model
 does not agree with the observations even if the parameter $\theta_c$
 is changed in a wide range. Figure \ref{fig:gaussian1d}
 shows the theoretical distribution of GRB isotropic energy
 for different star formation rate models. The theoretical distribution peaks at higher energy
 compared with the observed sample. Figure \ref{fig:gaussianks} shows the predicted and observed cumulative
 distributions of GRB isotropic energy for three different values of
 $L$, assuming the star formation rate model 2. Different star formation rate models and
 different values of $L$ only lead to a slight difference in the distribution.
 The Gaussian jet model predicts that most GRBs are detected
 within its central core where the isotropic energy is large and
 nearly constant. This is because the energy
 decreases exponentially at the wing and the GRB cannot be triggered by the
 detectors. However, we note that if we introduce some
 dispersion on the value of the central angle $\theta_c$, this
 inconsistency may be reduced (Zhang et al. 2004; Dai \& Zhang 2005).

\begin{figure}
\plotone{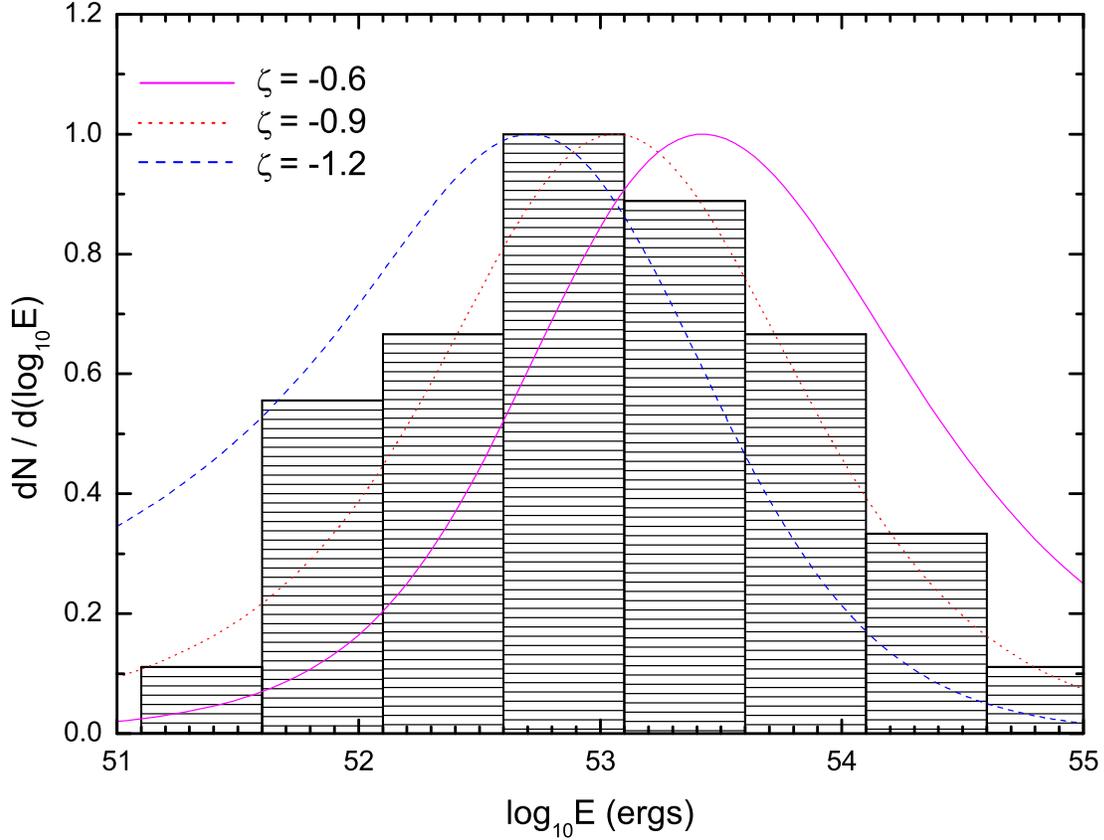} \caption{ Theoretical distribution of the
isotropic-equivalent energy in the uniform jet model in the case that the
distribution of isotropic-equivalent energy is a power-law function (lines),
compared with the observed distribution from a sample of 39 GRBs with known
fluences and redshifts detected so far (histogram). Different lines correspond
to different values of the power-law index $\zeta$. Here we take the star
formation rate model 2 and $L=-1$ as an example. \label{fig:ujpowerlaw1d}}
\end{figure}

\begin{figure}
\plotone{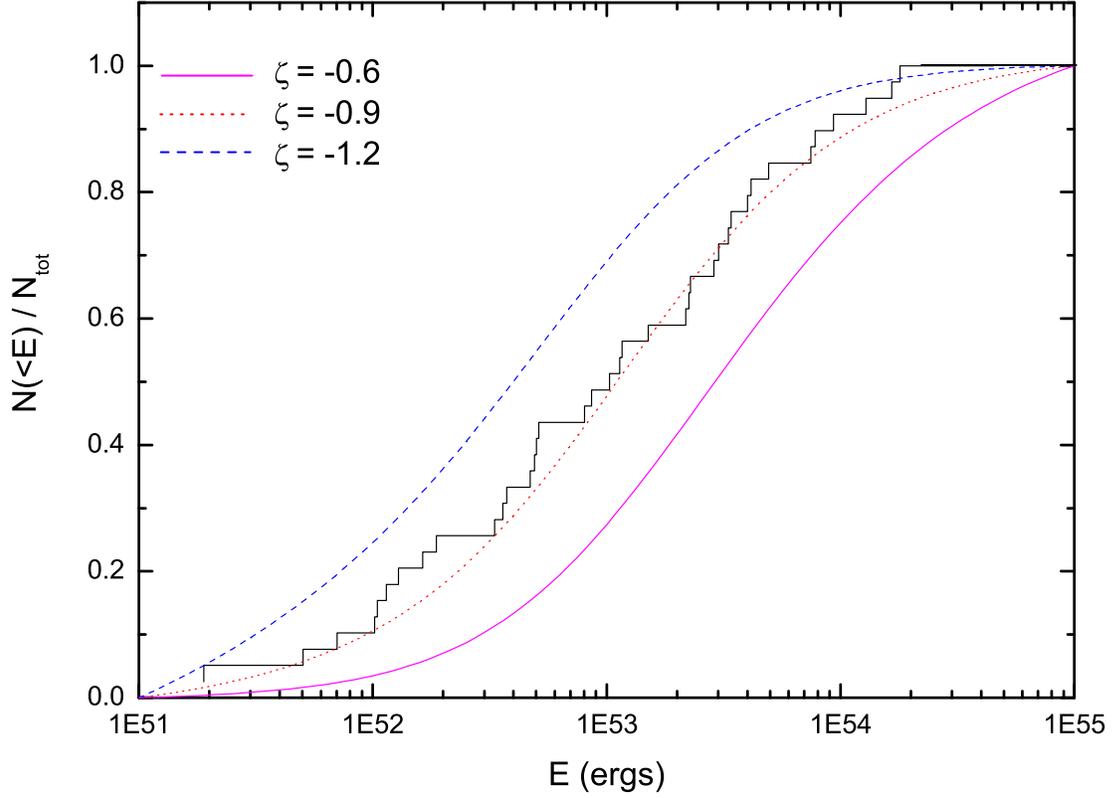} \caption{Cumulative distribution for the uniform jet model
in the case that the distribution of isotropic-equivalent energy is a power-law
function. Different lines correspond to the different values of the power-law
index $\zeta$. The cumulative histogram is plotted from the sample. Here we
take the star formation rate model 2 and $L=-1$ as an example.
\label{fig:ujpowerlawks}}
\end{figure}

 In the uniform jet model, we have considered three cases in which $f(E)$
 is a power-law, exponential, or Gaussian function. By comparing the theoretical calculation and
 the observational data using the K-S test, we obtain a confidence level for the hypothesis that the
 observational data is drawn from the theoretical distribution.
 In the case of a power-law distribution, the result of the K-S test depends mainly on the value
 of the index $\zeta$. Our numerical results show that when $\zeta$ varies between -0.99 and -0.79,
 the theoretical distribution is compatible with the observed data at the confidence level of $>40 \%$.
 This confidence reaches a maximum of $90 \%$ for $\zeta=-0.91$.
 We show in Figure \ref{fig:ujpowerlaw1d} the predicted and observed distributions of GRB isotropic
 energy for three different values of $\zeta$. Figure \ref{fig:ujpowerlawks} presents
 the predicted and observed cumulative distributions of GRB isotropic energy
 for three different values of $\zeta$.

\begin{figure}
\plotone{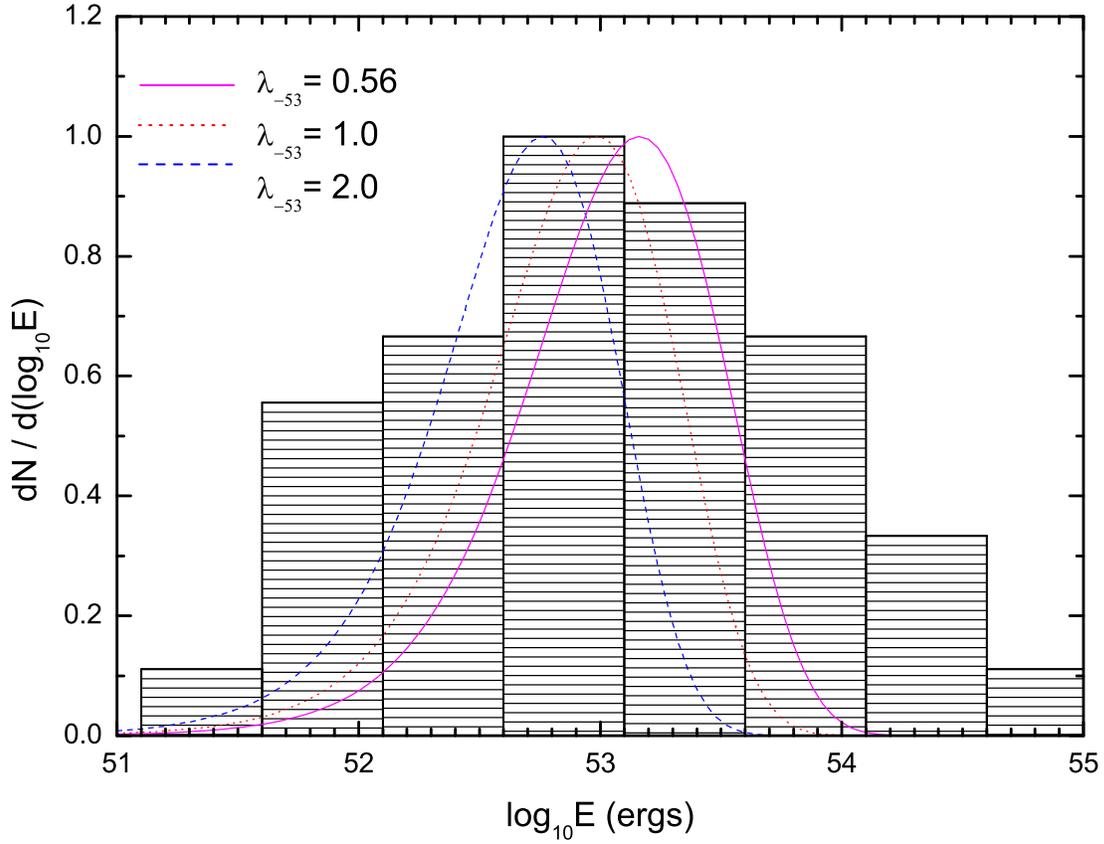} \caption{ Theoretical distribution of the
isotropic-equivalent energy in the uniform jet model in the case that the
distribution of isotropic-equivalent energy is an exponential function (lines),
compared with the observed distribution from a sample of 39 GRBs with known
fluences and redshifts detected so far (histogram). Different lines correspond
to different values of the  exponential parameter $\lambda$. Here we take the
star formation rate model 2 and $L=-1$ as an example. \label{fig:ujexp1d}}
\end{figure}

\begin{figure}
\plotone{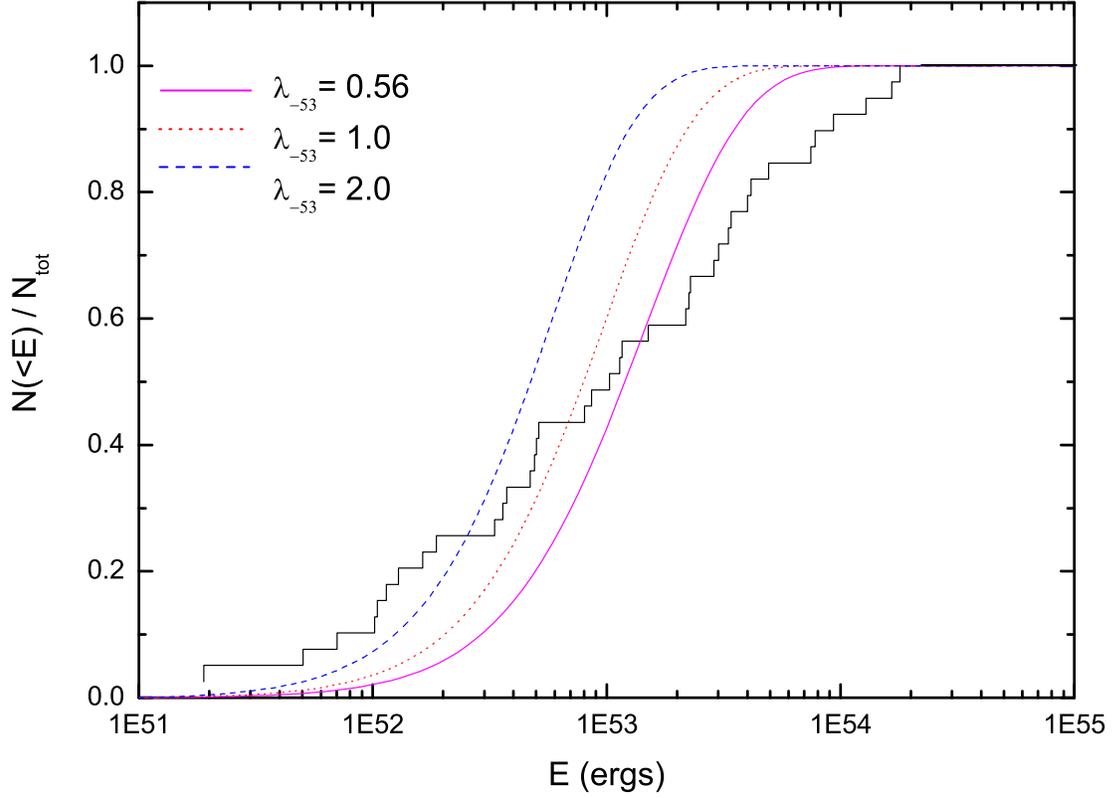} \caption{Cumulative distribution for the uniform jet model
in the case that the distribution of isotropic-equivalent energy is an
exponential function. Three values of $\lambda$ are considered. The cumulative
histogram is plotted from the sample. Here we take the star formation rate
model 2 and $L=-1$ as an example. \label{fig:ujexpks}}
\end{figure}

\begin{figure}
\plotone{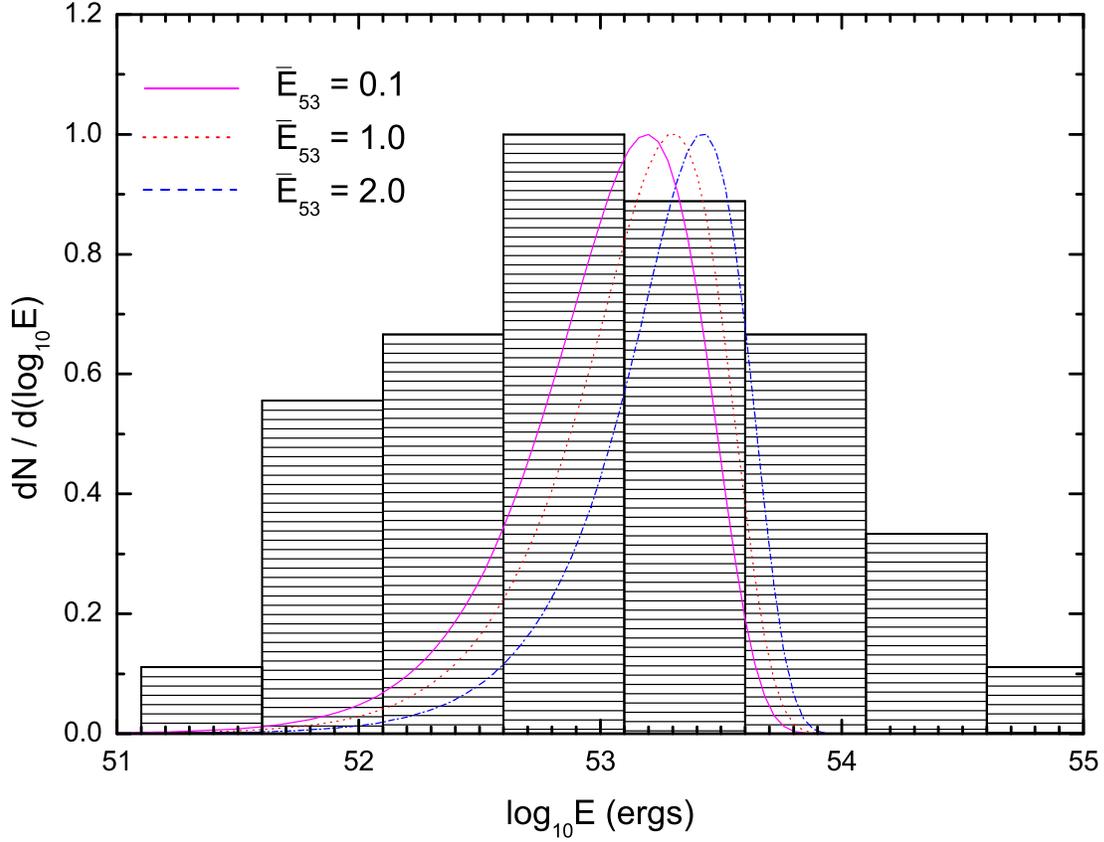} \caption{ Theoretical distribution of the
isotropic-equivalent energy in the uniform jet model in the case
that the distribution of isotropic-equivalent energy is a Gaussian
function (lines), compared with the observed distribution from a
sample of 39 GRBs with known fluences and redshifts detected so
far (histogram). Different lines correspond to different values of
mean energy $\bar{E}$ (in the units of ergs). We adopt the scatter
in the Gaussian function as $\sigma=1.7\times 10^{53}$ ergs. Here
we take the star formation rate model 2 and $L=-1$ as an
example.\label{fig:ujgaussian1d}}
\end{figure}

\begin{figure}
\plotone{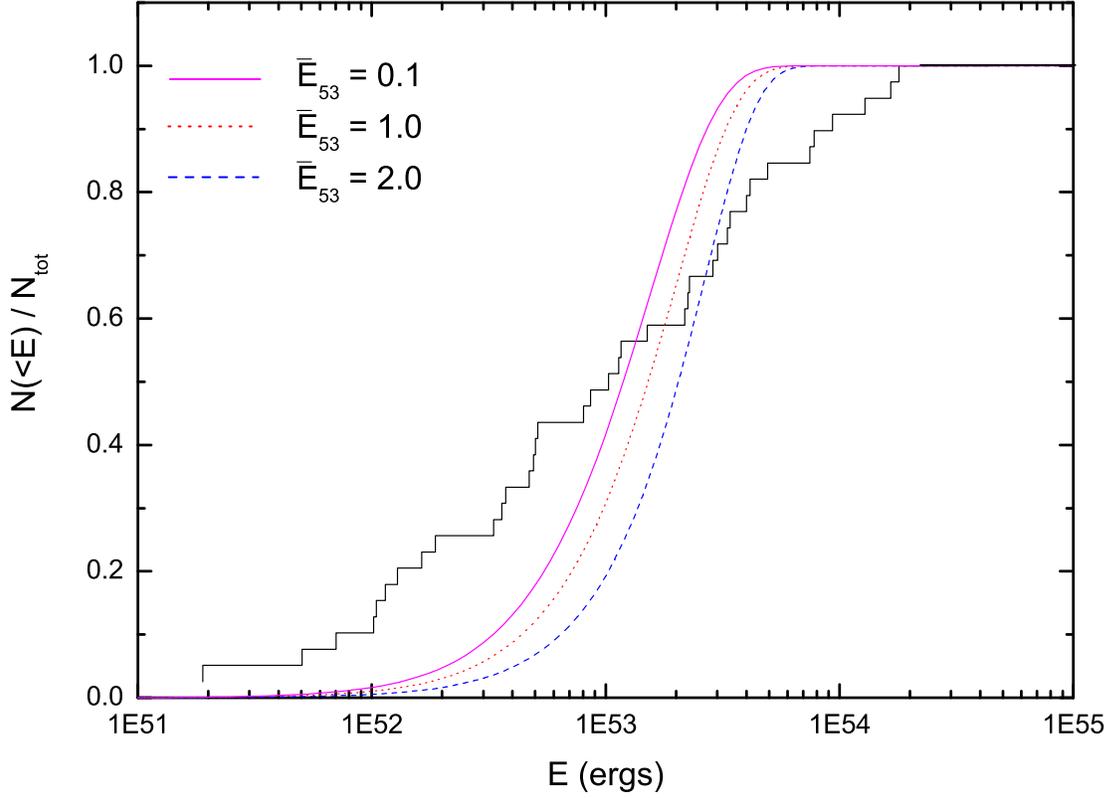} \caption{Cumulative distribution for the
uniform jet model in the case that the distribution of
isotropic-equivalent energy is a Gaussian function. Different
lines correspond to different values of mean energy $\bar{E}$ (in
the units of ergs). We adopt the scatter in the Gaussian function
as $\sigma=1.7\times 10^{53}$ ergs. The cumulative histogram is
plotted from the sample. Here we take the star formation rate
model 2 and $L=-1$ as an example. \label{fig:ujgaussianks}}
\end{figure}

 In the case of exponential and Gaussian distributions,
 the results of the K-S test also depend on the parameters. But
 the theoretical distributions are consistent with the observations
 at the confidence level of $<5 \%$, even though the parameters are changed
 greatly. Figures \ref{fig:ujexp1d} and \ref{fig:ujexpks} show
 the differential and cumulative distributions of GRB isotropic energy
 respectively. Despite the variation of the parameter $\lambda$,
 the figures show a poor consistency between the theoretical prediction and
 the observed data. Figures \ref{fig:ujgaussian1d} and
 \ref{fig:ujgaussianks} exhibit similar results for the case that $f(E)$
 follows a Gaussian function.

 We next consider the 2D distribution
 $\dot{N}(E,z)$. The distribution of GRBs is a function of both isotropic
 equivalent energy and redshift. In the
 1D analysis, the information contained in the redshift space is
 concealed by the integration over redshift. In order to explore
 the overall information about the distribution of GRBs, we extend
 the distribution from one dimension to two dimensions, and compare it
 with the observed sample (see Figs. 12-16). We performed 2D K-S test
 to check the consistency of the predicted 2D distribution with the observations.
 For the sample of 39 bursts, a 2D K-S test shows that for the power-law structured jet model,
 the predicted distribution is consistent with the observations at the confidence level of
 $>40\%$ while the parameters $k_E$ and $\theta_c$ are changed in a wide
 range. When adopting the value of parameters $k_E=2.2$ and $\theta_c=0.02$,
 we obtain the best fit, at the confidence level of $58\%$.
 For the uniform jet model in the power-law case, the model with SFR2, $L=-1$,
 and $\zeta=-0.92$ corresponds to a best fit, at the confidence level of
 $57\%$. When $\zeta$ varies between -0.97 and -0.85, the predicted distribution is
 compatible with the observed data at the confidence level of $>40 \%$.
 In the exponential and Gaussian case, we found that the
 hypothesis that the data are drawn from these three models is completely rejected.
 This conclusion is unchanged for different values of the parameters.

 However, this conclusion is tentative, because the sample suffers from
 various observational biases, such as GRB detection, localization, and
 especially the selection effect in the identification of redshift (Bloom 2003).
 Nakar et al. (2004) suggested that the selection effect in
 redshift can be minimized by testing the $\theta$ distribution for a given
 redshift. But the size of the sample is greatly reduced in this
 method. With the cumulation of the sample in the future, this
 method might be used to test the configuration of jets in
 GRBs.

\section{Discussion and Conclusion}
\label{sec:discussion}
In this paper, we have worked out the theoretical distribution of
burst isotropic-equivalent energy in 1D $\dot{N}(E)=dN/dE$ and in
2D $\dot{N}(E,z)=dN^2/dEdz$ for several jet structure models. The
figures and analyses above show that theoretically the GRB rates
predicted by different models differ from each other greatly. It
is possible for us to find the true model by the method shown here
if the sample is large enough. Based on the theoretical analysis
in \S \ref{sec:numerical-ana}, we carried out numerical
calculations and compared $\dot{N}(E)=dN/dE$ and
$\dot{N}(E,z)=dN^2/dEdz$ with the the observed sample of GRBs with
known fluences and redshifts. Via K-S test, we found that the
power-law structured jet model and the uniform jet model can be
consistent with the currently observed data at a confidence level
of $\geqslant40\%$ under certain circumstances.

One advantage of our work over Perna et al. (2003) and Nakar et al.
(2004) lies in the fact that the sample of bursts with measured
redshifts and fluences is larger than the one with known redshifts
and modelled angles. The sample of bursts with known redshift and
modelled angle is too small (currently 16 GRBs). According to the
relation $t_{j}\propto \theta_{j}^{8/3}$ (Sari et al. 1999), if the
jet opening angle $\theta_{j}$ is small, the jet break time $t_{j}$
probably is so early that there is only the upper limit on it. In
the contrary, if $\theta_j$ is large, the late-time optical
transient may be too dim to be detected. As a result, this selection
effect in detecting $t_{j}$ leads to a small sample of bursts with
measured jet break times. However, the sample of bursts with
measured fluences and redshifts does not suffer from such selection
effect. This larger sample helps us to draw a stronger conclusion.

However, it is worth noting that all results above are obtained
under the following assumptions. First, the isotropic-equivalent
energy release in gamma-ray band is assumed to be proportional to
the total isotropic-equivalent energy of the bursts. Second, we
consider the simplifying assumption that the theoretical model is
strictly consistent with the actual jet structure. These
theoretical models predict a smooth, broken power-law light curve.
However, recent observations revealed variabilities in some
afterglow light curves such as in GRB $021004$ (Fox et al. 2003).
One leading model named ``patchy shell" model attributes this
fluctuation in light curves to random angular fluctuation of the
energy per solid angle (Kumar \& Piran 2000; Nakar, Piran, \&
Granot 2003). In this paper, we consider a regular energy
distribution in the jet rather than with some random angular
fluctuations. This simplifying assumption may result in an
underestimate of the number of energetic events.

Third, the current sample of 39 bursts with measured redshifts
suffers from numerous selection effects, especially in optical
afterglow detections and redshift identifications. To reduce these
effects, we used a strategy suggested by Bloom (2003). On the
other hand this method might introduce a systematic error into the
result. To reach a robust conclusion, Nakar et al. (2004) put
forward an approach to minimize the selection effect in $z$, i.e.,
to test the distribution for a given z. In our work, to avoid this
selection effect and reduce the dependence on SFR model, we need
to derive the distribution of isotropic-equivalent energy for a
fixed redshift $\dot{N}(E)|_{z}$. However, this method requires a
large sample of bursts with measured fluences and with the same
redshift, while there are only a few bursts are detected with the
same redshift. With a larger sample this method shown here can be
expected to give a stronger conclusion. With the advent of the
{\em Swift} era, the sample of GRBs with measured fluence and
redshift is anticipated to be much larger than the current one.
The large sample of bursts detected by the same detector will be
having the same parameters such as the detector sensitivity, thus
reducing the uncertainty and reaching a stronger conclusion.

\acknowledgments We thank E. W. Liang, D. Xu and Z. P. Jin for
their helpful discussions. This work was supported by the National
Natural Science Foundation of China (grants 10233010 and 10221001)
and the Ministry of Science and Technology of China (NKBRSF
G19990754).

\clearpage

\textwidth=6.9in \textheight=9.0in
\columnsep=5.0mm 
\parindent=6.0mm
\voffset=-5mm \hoffset=-5mm

\begin{deluxetable}{llccccl}
\rotate \tabletypesize{\scriptsize} \tablewidth{6.2in}
\tablecaption{Compilation of Spectra and Energetics Input
Data\label{table1}} \tablecolumns{11} \tablehead{
\colhead{GRB{\tablenotemark{a}}}        & \colhead{$z$}
&\colhead{$S_{\gamma}${\tablenotemark{b}}} &\colhead{Bandpass}
&\colhead{$[\alpha, \beta]\tablenotemark{c}$}  & \colhead{$E^{\rm
obs}_p${\tablenotemark{~d}}}
        & \colhead{References}\\
\colhead{}          & \colhead{} &\colhead{[$10^{-6}$ {\tiny erg
cm$^{-2}$}]}    &\colhead{[keV]} &\colhead{} & \colhead{[kev]}  &
\colhead{($z$, $S_{\gamma}=S$, $t_{\rm jet}=t$, $n$, $\alpha$,
$\beta$, $E_p$)}} \startdata
970228  &        0.6950         &        11.00    &        40, 700        &            -1.54, -2.50    &        115        &               $z$: 1, $S$: 2, $\alpha$: 2, $\beta$: 2, $E_p$: 2    \\
970508  &        0.8349         &         1.80    &        40, 700        &           -1.71, -2.20    &        79         &               $z$: 3, $S$: 2,  $\alpha$: 2, $\beta$: 2, $E_p$: 2    \\
970828  &        0.9578         &        96.00        &        20, 2000       &            -0.70, -2.07    &        298        &               $z$: 4, $S$: 5, $\alpha$: 6, $\beta$: 5, $E_p$: 5    \\
971214  &        3.4180         &         8.80    &        40, 700        &               -0.76, -2.70    &        155        &               $z$: 7, $S$: 2, $\alpha$: 2, $\beta$: 2, $E_p$: 2    \\
980326  &        1.0000         &        0.75    &        40, 700        &            -1.23, -2.48    &        47          &             $z$: 8, $S$: 2, $\alpha$: 2, $\beta$: 2, $E_p$: 9  \\
980425  &        0.0085         &         3.87       &        20, 2000       &                -1.27, -2.30  $*$    &        118        &            $z$: 10, $S$: 5, $\alpha$: 5, $E_p$: 5       \\
980613  &        1.0969         &         1.00    &        40, 700        &              -1.43, -2.70    &        93        &             $z$: 11, $S$: 2, $\alpha$: 2, $\beta$: 2, $E_p$: 2  \\
980703  &        0.9662         &        22.60         &        20, 2000       &              -1.31, -2.40    &        254        &              $z$: 12, $S$: 5, $\alpha$: 6, $\beta$: 5, $E_p$: 5         \\
990123  &        1.6004         &        300.00     &        40, 700        &               -0.89, -2.45    &        781        &               $z$: 13, $S$: 2, $\alpha$: 2, $\beta$: 2, $E_p$: 2  \\
990506  &        1.3066         &        194.00     &        20, 2000            &        -1.37, -2.15    &        283        &              $z$: 14, $S$: 5, $\alpha$: 6, $\beta$: 5, $E_p$: 5   \\
990510  &        1.6187         &        19.00    &        40, 700        &                -1.23, -2.70     &        163        &         $z$: 15, $S$: 2,  $\alpha$: 2, $\beta$: 2, $E_p$: 2         \\
990705  &        0.8424         &        75.00    &        40, 700        &               -1.05, -2.20     &        189        &             $z$: 16, $S$: 2,  $\alpha$: 2, $\beta$: 2, $E_p$: 2  \\
990712  &        0.4331         &        11.00    &        2, 700         &                -1.88, -2.48     &        65         &              $z$: 15, $S$: 17, $\alpha$: 2, $\beta$: 2, $E_p$: 2         \\
991208  &        0.7055         &        100.00      &        25, 1000       &               \nodata, \nodata        &        \nodata        &            $z$: 18, $S$: 19  \\
991216  &        1.0200         &        194.00   &        20, 2000       &                -1.23, -2.18    &        318       &           $z$: 20, $S$: 5,  $\alpha$: 6, $\beta$: 5, $E_p$: 5         \\
000131  &        4.5000         &        35.10    &        26, 1800       &              -1.20, -2.40     &        163       &             $z$: 21, $S$: 21,  $\alpha$: 21, $\beta$: 21, $E_p$: 21      \\
000210  &        0.8463         &        61.00    &        40, 700        &                \nodata, \nodata        &        \nodata        &           $z$: 22, $S$: 22 \\
000214  &        0.4200         &         1.42    &        40, 700        &               -1.62, -2.10     &        $>$ 82         &            $z$: 23, $S$: 2, $\alpha$: 2, $\beta$: 2, $E_p$: 2   \\
000301c &        2.0335         &         2.00    &        150, 1000      &                \nodata, \nodata        &        \nodata        &          $z$: 24, $S$: 25,   \\
000418  &        1.1182         &        20.00     &        15, 1000       &                 \nodata, \nodata        &        \nodata        &              $z$: 14, $S$: 26  \\
000911  &        1.0585         &        230.00     &        15, 8000       &               -1.11, -2.32    &        579       &             $z$: 27, $S$: 27, $\alpha$: 27, $\beta$: 27, $E_p$: 27      \\
000926  &        2.0369         &         6.20        &        25, 100        &              \nodata, \nodata        &        \nodata        &              $z$: 28, $S$: 29  \\
010222  &        1.4769         &        120.00   &        2, 700         &                -1.35, -1.64     &        $>$ 358        &            $z$: 30, $S$: 31,  $\alpha$: 2, $\beta$: 2, $E_p$: 2        \\
010921  &        0.4509         &        18.42   &        2, 400              &        -1.55, -2.30     &        89         &           $z$: 32, $S$: 33, $\alpha$: 33, $\beta$: 34, $E_p$: 33      \\
011121  &        0.3620         &        24.00       &        25, 100        &             -1.42, -2.30 $*$        &        217        &             $z$: 35, $S$: 36,  $\alpha$: 6, $E_p$: 6     \\
011211  &        2.1400         &         5.00      &        40, 700        &                 -0.84, -2.30  $*$      &        59          &            $z$: 37, $S$: 37, $\alpha$: 6, $E_p$: 6     \\
020124  &        3.1980         &         8.10  &        2, 400            &                  -0.79, -2.30       &        120         &              $z$: 38, $S$: 33, $\alpha$: 33, $\beta$: 34, $E_p$: 33      \\
020405  &        0.6899         &        74.00        &        15, 2000       &                 0.00, -1.87     &       192        &            $z$: 39, $S$: 39,  $\alpha$: 39, $\beta$: 39, $E_p$: 40      \\
020813  &        1.2550         &        102.00     &        30, 400         &               -1.05, -2.30    &        212       &              $z$: 41, $S$: 42, $\alpha$: 42, $\beta$: 42, $E_p$: 42      \\
021004  &        2.3351         &         2.55      &        2, 400         &                  -1.01, -2.30  $*$       &        80           &     $z$: 43, $S$: 33,  $\alpha$: 33, $E_p$: 33     \\
021211  &        1.0060         &         3.53      &        2, 400         &                 -0.80, -2.37     &        47         &              $z$: 44, $S$: 45,  $\alpha$: 45, $\beta$: 45, $E_p$: 45      \\
030226  &        1.9860         &         5.61      &        2, 400         &                 -0.89, -2.30   &        97         &              $z$: 46, $S$: 45,  $\alpha$: 45, $\beta$: 34, $E_p$: 45      \\
030323  &        3.3718         &         1.23     &        2, 400         &               -1.62, -2.30  $*$       &        \nodata        &              $z$: 47, $S$: 33, $\alpha$: 33      \\
030328  &        1.5200         &        36.95      &        2, 400         &                  -1.14, -2.09     &        126        &            $z$: 48, $S$: 33, $\alpha$: 33, $\beta$: 33, $E_p$: 33      \\
030329  &        0.1685         &        110.00     &        30, 400         &                 -1.26, -2.28     &        68          &             $z$: 49, $S$: 50, $\alpha$: 50, $\beta$: 50, $E_p$: 50     \\
031203  &        0.1055         &         1.20       &        20, 2000       &              -1.00 $*$, -2.30    $*$     &        $>$ 190        &              $z$: 51, $S$: 52, $E_p$: 53  \\
040924  &        0.8590         &         2.73    &        20, 500        &                -1.17, -2.30  $*$      &        67          &           $z$: 54, $S$: 55,  $\alpha$: 56, $E_p$: 55   \\
041006  &        0.7160         &         19.90     &        25, 100        &              -1.37, -2.30   $*$     &        63          &              $z$: 57, $S$: 58, $\alpha$: 56, $E_p$: 56   \\
050408  &        1.2357         &       1.90        &      30, 400          &             -1.98, -2.30    $*$       &        20        &       $z$: 59, $S$: 59, $\alpha$: 60, $E_p$: 60\\
050525a &        0.6060         &       20.00       &  15, 350               &        -1.00, -2.30     $*$       &       79               &   $z$: 61, $S$: 61, $\alpha$: 61, $E_p$: 61\\
050603  &        2.2810         &       34.10       &  20, 3000              &      -0.79, -2.15             &      349             &     $z$: 62, $S$: 63, $\alpha$: 63, $\beta$: 63, $E_p$: 63  \\

\enddata

\tablenotetext{a} {Upper/lower limit data are indicated with $<$
and $>$ respectively. References are given in order for redshift
(``$z$''), fluence (``$S$''), low energy band spectral slope
(``$\alpha$''), high energy band spectral slope (``$\beta$''), and
spectral peak energy (``$E_p$'').}

\tablenotetext{b} {GRB fluence $S_{\gamma}$ calculated in the
observed bandpass [$e_1$, $e_2$] keV.}

\tablenotetext{c} {\ Low energy ``Band'' spectral slope $\alpha$
and high energy ``Band'' spectral slope $\beta$. When $\beta$ is
reported in the literature but $\alpha$ is not, we set
$\alpha=-1.00$ (marked with $*$). Following Atteia (2003), when
$\alpha$ is reported in the literature and $\beta$ is not, we fix
$\beta$=-2.30 (marked with $*$). When both $\alpha$ and $\beta$
are not reported in the literature we assume $\alpha$=-1.00 and
$\beta$=-2.30. }

\tablenotetext{d} { Observed spectral peak energy $E^{\rm obs}_p =
E^{\rm obs}_o(2+\alpha)$. When $E^{\rm obs}_p$ is not reported we
fix $E^{\rm obs}_p=250$ keV. When $E^{\rm obs}_p$ is reported only
with a limit, we adopt $E^{\rm obs}_p$ as the value of the limit.}

\tablerefs{ \tiny 1. Bloom et al. 2001a; 2. amati et al. 2002; 3.
Bloom et al. 1998; 4. Djorgovski et al. 2001; 5. Jimenez et al.
2001; 6. Amati 2004; 7. Kulkarni et al. 1998; 8. Bloom et al.
1999; 9. Groot et al. 1998; 10. Tinney et al. 1998; 11.
Djorgovskei et al. 2003; 12. Djorgovski et al. 1998; 13. Kulkarni
et al. 1999; 14. Bloom et al. 2003; 15. Vreeswijk et al. 2001; 16.
Le Floc'h et al. 2002; 17. Frontera et al. 2001; 18. Djorgovski et
al. 1999a; 19. Hurley et al. 2000; 20. Vreeswijk et al. 1999;
Djorgovski et al. 1999b; 21. Andersen et al. 2000; 22. Piro et al.
2002; 23. Antonelli et al. 2000; 24. Castro et al. 2000a; 25.
Jensen et al. 2001; 26. Berger et al. 2001; 27. Price et al.
2002c; 28. Castro et al. 2000b; 29. Price et al. 2001; 30. Mirabal
et al. 2002; 31. in't Zand et al. 2001; 32. Price et al. 2002a;
33. Sakamoto et al. 2004; 34. Atteia 2003; 35. Garnavich et al.
2003; 36. Price et al. 2002b ; 37. Holland et al. 2002; 38. Hjorth
et al. 2003; 39. Price et al. 2003a ; 40. Price et al. 2003b; 41.
Barth et al. 2003; 42. Barraud et al. 2003; 43. M\"{o}ller et al.
2002; 44. Vreeswijk et al. 2003; 45.Crew et al. 2003; 46. Greiner
et al. 2003; 47. Vreeswijk et al. 2004; 48. Martini et al. 2003;
49. Bloom et al. 2003d; 50. Vanderspek 2004a; 51. Prochaska et al.
2003; 52. Watson et al. 2004; 53. Sazonov et al. 2004; 54.
Wiersema et al. 2004; 55. Golenetskii et al. 2004; 56. Vanderspek
2004b; 57. Price et al. 2004; 58. Galassi et al. 2004; 59. Berger
et al. 2005; 60. Sakamoto et al. 2005; 61. Cummings et al. 2005;
62. Berger et al. 2005; 63. Golenetskii et al. 2005.}
\end{deluxetable}

\begin{figure}
\plotone{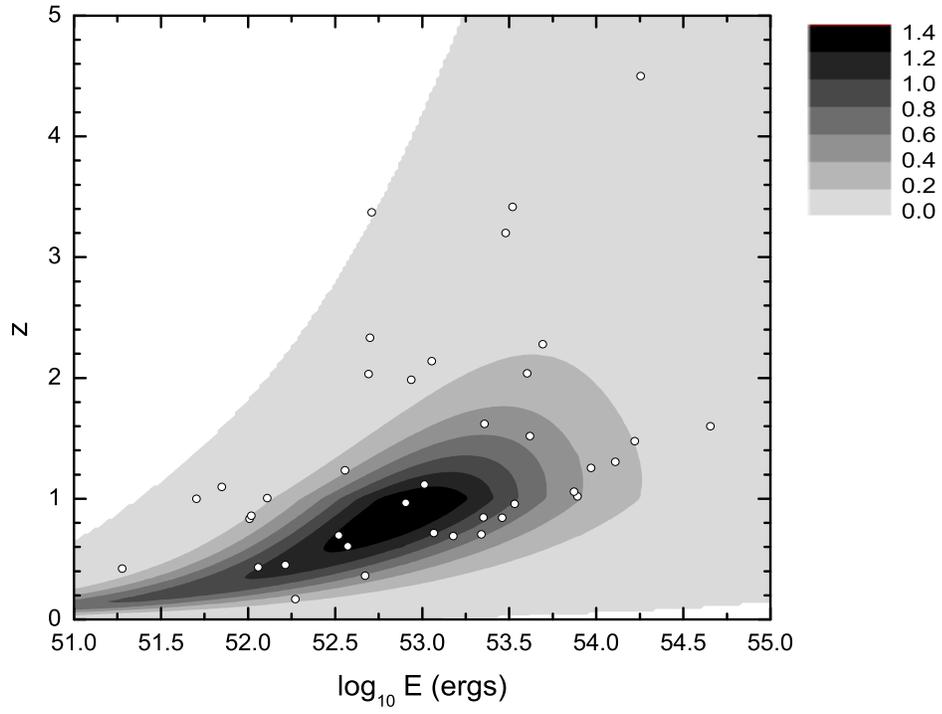} \caption{The 2D distribution density,
$\dot{N}(E,z)$, for the power-law structured jet model and star
formation rate model 2. The parameters are assumed: $L=-1$,
$k_E=2.2$, and $\theta_c=0.02$. The circles denote 39 bursts with
known fluences and redshifts detected so far.
\label{fig:gaussiancontour}}
\end{figure}

\begin{figure}
\plotone{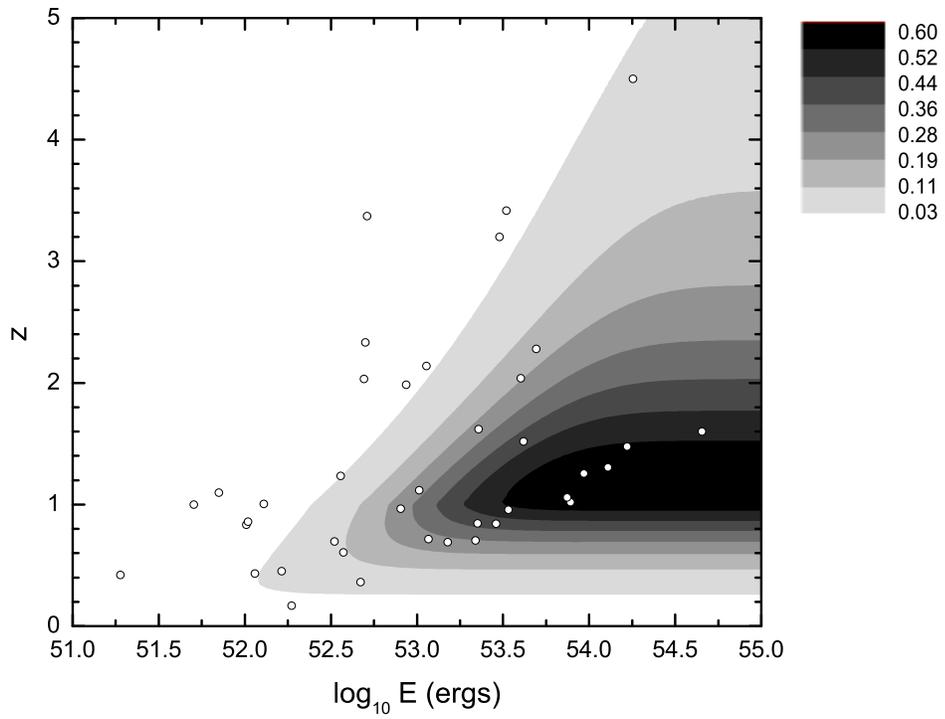} \caption{The 2D distribution density, $\dot{N}(E,z)$, for
the Gaussian structured jet model, star formation rate model 2. The parameters
are assumed: $L=-1$ and $\theta_c=0.05$. The circles denote 39 bursts with
known fluences and redshifts detected so far. \label{fig:powerlawcontour}}
\end{figure}

\begin{figure}
\plotone{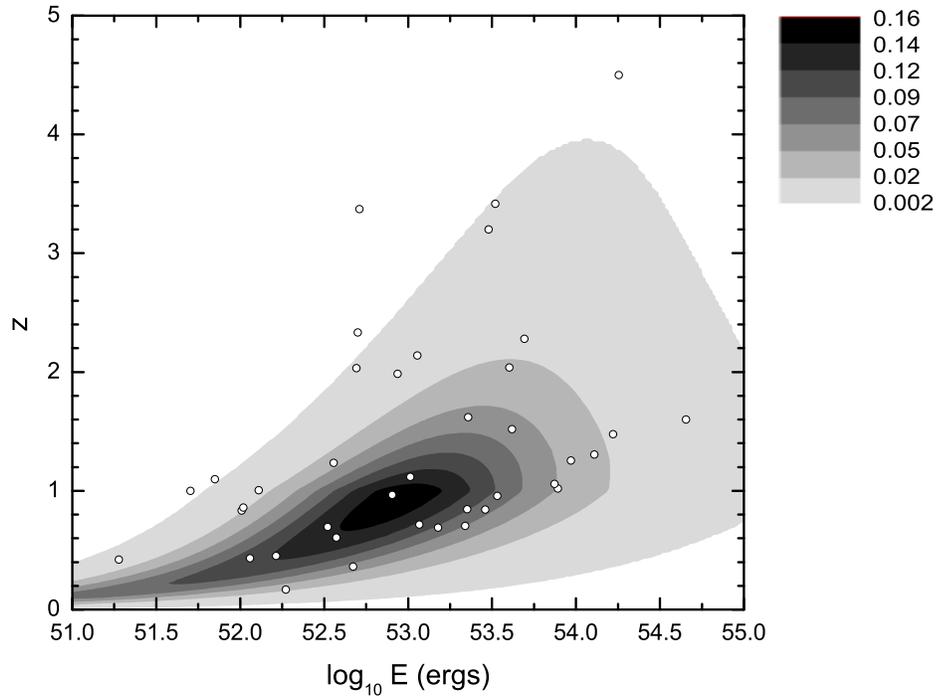} \caption{The 2D distribution density,
$\dot{N}(E,z)$, for the uniform jet model (the power-law case) and
star formation rate model 2. The parameters are assumed: $L=-1$
and $\zeta=-0.9$. The circles denote 39 bursts with known fluences
and redshifts detected so far. \label{fig:ujpowerlawcontour}}
\end{figure}

\begin{figure}
\plotone{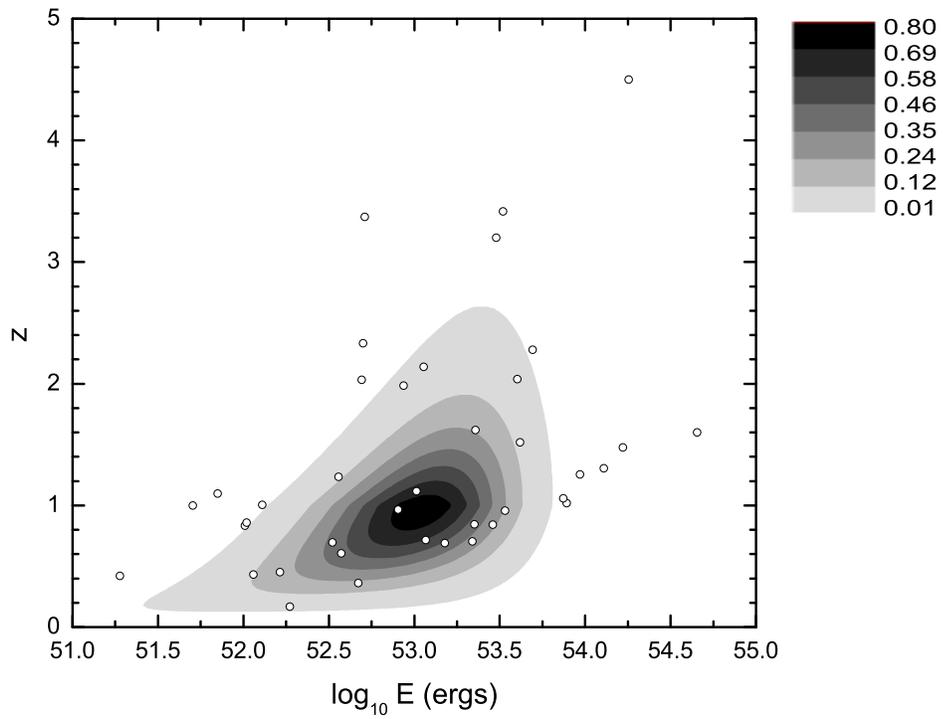} \caption{The 2D distribution density, $\dot{N}(E,z)$, for
the uniform jet model (the exponential case) and star formation rate model 2.
The parameters are assumed: $L=-1$ and $\lambda=0.7\times 10^{-53}~\rm
{ergs^{-1}}$. The circles denote 39 bursts with known fluences and redshifts
detected so far.\label{fig:ujexpcontour}}
\end{figure}

\begin{figure}
\plotone{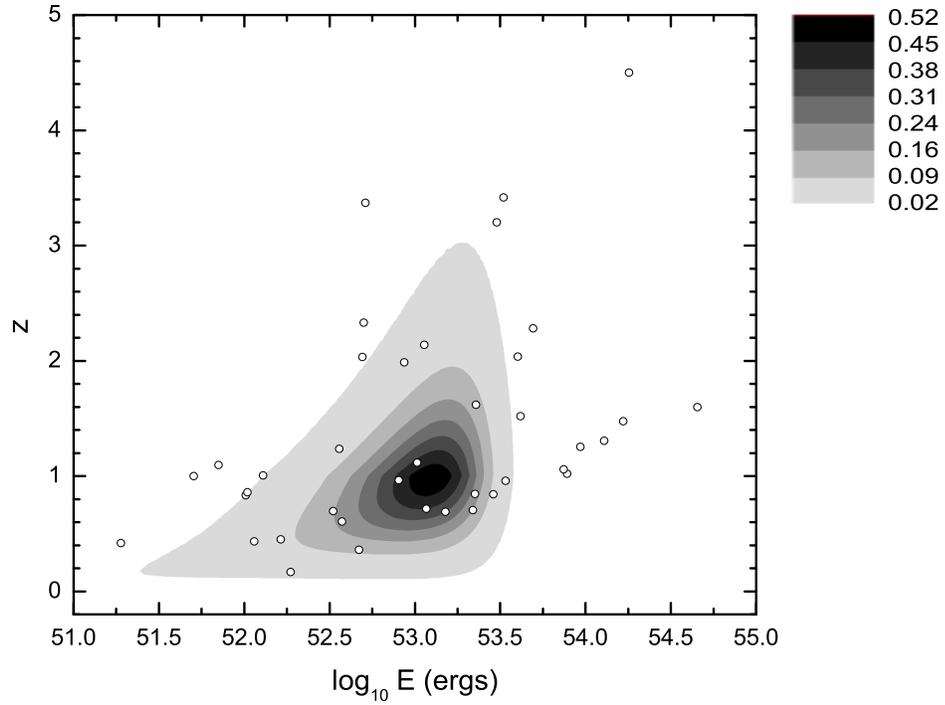} \caption{The 2D distribution density,
$\dot{N}(E,z)$, for the uniform jet model (the Gaussian case) and
star formation rate model 2. The parameters are assumed: $L=-1$
and $\bar{E}=0.7\times 10^{53}$ ergs, and $\sigma=1.0\times
10^{53}$ ergs. The circles denote 39 bursts with known fluences
and redshifts detected so far. \label{fig:ujgaussiancontour}}
\end{figure}

\end{document}